# Open-domain question classification and completion in conversational information search

Omid Mohammadi Kia
*Faculty of Computer Science and Engineering*
*Shahid Beheshti University*
Tehran,Iran
o.mohammadikia@mail.sbu.ac.ir

Mahmood Neshati
*Faculty of Computer Science and Engineering*
*Shahid Beheshti University*
Tehran,Iran
m_neshati@sbu.ac.ir

Mahsa Soudi Alamdari
*Faculty of New Sciences and Technologies*
*Tehran University*
Tehran,Iran
ms.alamdari@ut.ac.ir

*Abstract*—Searching for new information requires talking to the system. In this research, an Open-domain Conversational information search system has been developed. This system has been implemented using the TREC CAsT 2019 track, which is one of the first attempts to build a framework in this area. According to the user's previous questions, the system firstly completes the question (using the first and the previous question in each turn) and then classifies it (based on the question words). This system extracts the related answers according to the rules of each question. In this research, a simple yet effective method with high performance has been used, which on average, extracts 20% more relevant results than the baseline.

## I. Introduction

In the past, the search was being done in such a way that the user entered the request in text, and in response to that, a limited number of links related to the request were displayed to the user. Today, the shrinking of the devices used for search, as well as advances in the field of automatic speech recognition (ASR), have led to huge changes in the way the user and device interact. The way of interaction is no longer limited to choosing a link, but it is a collaboration between the user and the device. This has led to the popularity of text-based smart assistants as well as voice-based assistants (such as Cortana or Siri) around the world. Currently, over 50% of Internet users browse the Internet through their mobile phones [1]. According to research [2], 40% of searches are done by voice. In recent years, conversational search engines have received a great deal of attention from Information Retrieval (IR) and Natural Language Processing (NLP) communities [3]. Users need to ask several questions about a topic because in many cases they do not know what they are looking for or cannot express it.

The focus of this research is on text-based conversational search. In this type of search, users can ask several consecutive questions about a topic. These questions are asked for more information, and each question is usually explicitly or implicitly related to the previous questions.

Conversational Search (CS) differs from both IR and Question Answering (QA). The focus of IR systems is on document retrieval from a large set of documents. If the user requires complex information, searching for results will be difficult and inefficient. Also, IR generally does not process the user's language and almost does word matching. In QA, researchers seek to find effective answers to user's special questions. QA is involved in understanding the meaning of the questions [4]. However, this does not mean constant interaction to match results, nor does it mean asking system questions to the user. CS, on the other hand, involves the constant exchange of information between the server and the user. The ability to learn and mutual benefits in IR and QA methods can be a part of the CS approach.

Bidirectional Encoder Representations from Transformers (BERT) has been used in most recent methods [5], [6] to increase accuracy. However, because such methods process each pair (query - documents) along with a vast neural network, they dramatically increase the cost of calculating the query-document relevance score [8], [9]. According to Article [9], the computational time required by BERT is about 10000 milliseconds, but the time of methods such as BM25 is only 100 milliseconds. The purpose of this study is to provide a way to increase the accuracy of search engines without using neural networks. In this way, in addition to maintaining the accuracy of the results, the computation time remains 100 milliseconds.

Due to the time and computational complexity of existing neural networks, these networks usually perform analysis on a limited set of documents, which poses many challenges. The best initial information retrieval methods are performed by the Query Likelihood (QL) with pseudo-relevance feedback RM3 [10], [11] or BM25 + RM3. These methods, especially in TREC CAsT, where the user request is short, do not provide promising results with the Recall criterion and This is a major challenge for neural network-based models [12]. In this research, a rule-based method is presented which does not have the mentioned problems with the neural networks.

The basis of this approach is to develop user requests and classify questions to increase the effectiveness of the results. It should be noted that the results of the conversation and code analysis have been published publicly[1].

In the following, Section 2 reviews the previous works related to this article. At the beginning of Section 3, the problem is formally defined. Then, two main steps in implementing the proposed methods are explained. Section 4 examines the data set, implementation, and hyperparameters. Finally, in Section 6, the conclusion is announced and the plans for the future are presented.

## II. PREVIOUS WORKS

Conversations can be divided into three general categories in terms of application: 1- Conversation for entertainment [13] 2- Conversation for task completion [14] (such as booking a ticket or ordering food) 3- Conversation for gathering information [15]. In this research, the last case is investigated. Previously, most of the studies were related to the first two categories and less attention was paid to the third type of conversations [16].

In terms of the response method, there are two types of methods: 1- Generating the answer by the system [17] 2- Retrieving the answer from the available answers [15]. The first type usually produces answers that are general and lack specific knowledge for the user, besides, the sentences may not be clear enough for the user. In the case of the second type, although the disadvantages of the previous type does not exist, the answers may be long. Given that CS is commonly used on devices with small screens, this length of response is a nuisance. In the article [16], the neural network has been used, which has led to an increase in the accuracy of the results, but the long time spent in this method is one of its major disadvantages. Two important features of the article method include the following:

1. In the conversation with the user, the user expresses the request very briefly, so first the system finds some of the top documents related to the user's request and gives both the documents and the user's request as input to the system to get better results.

2. Using the questions and answers in the dataset, other words related to the query words are extracted.

In conventional IR methods, the focus is solely on word similarities. However, in QA methods, there may not be a common word between question and answer (For example, one person asked, "My Windows is constantly shutting down," and another replied, "Because you did not completely delete your antivirus.") and the processing is based on the meaning of the words. This article [16] has enriched the answers by using the questions and answers information available in the knowledge base. TREC CAsT data have shown that neural networks cannot extract the text background well and weaken the results compared to the median of runs [15]; Therefore, in this study, a rule-based method is used, which, unlike the neural network, requires little processing load and time.

The classification of questions alone has been extensively studied [18] - [21]. The article [22] is about the QA in the field of medicine and the type of question is determined by the words, but how to find the relevant answers has not been studied. The article discusses only one question individually and does not take into account the impact of previous questions. It should be noted that our method is not limited to the medical domain and can be used in any field.

## III. THE PROPOSED METHOD

In this section, first the problem is formally defined, then our proposed method is described in three steps. In the first step, the required pre-processing is done on the text, then in the second step, the type of question is identified and the relevant words are added to the user's request. Finally, in the third step, the user's previous questions are added to the user request.

### A. Method Definition

The TREC CAsT contains the following data sets for information retrieval:

$$D = \{(Q_i, R_i, Y_i)\}_{i=1}^{N} \qquad (1)$$

In (1), $Q_i = \{q_i^1, q_i^2, q_i^3, \dots, q_i^{t-1}, q_i^t\}$. $\{q_i^1, q_i^2, q_i^3, \dots, q_i^{t-1}\}$ are the previous questions and the background of the user's session with the system, and $q_i^t$ is the user's current question of the system. $R_i = \{r_i^1, r_i^2, r_i^3, \dots, r_i^{m-1}, r_i^m\}_{m=1}^{M}$ are the answers to each user question. $Y_i = \{y_i^1, y_i^2, y_i^3, \dots, y_i^{m-1}, y_i^m\}_{m=1}^{M}$ is the degree to which each document relates to each question. In the above relationships, $N$ is the number of different user sessions with the system, $t$ is the number of user questions per session, and $M$ is the number of documents retrieved by the system. $y_i^m$ can take the values 0,1 and 2. The number two indicates the highest degree of relevance, it means, if $y_i^k = 2$, the document $r_i^k$ is related to the question $Q_i$.

### B. Text Processing

Some documents in the dataset were duplicates that were removed from the collection in the beginning. For each document, the stemming is performed, and stop words that do not change the content are removed. Abbreviations are replaced with full words, and pronoun references are modified using the AllenNLP tool [23], [24].

### C. Classification of questions and type of the answer of each category

Most questions can be classified by short structures according to the first word and the keywords of the question. Most questions can be divided into structures such as (Who…?), (Where…?) And (When…?). After determining the type of question, the answers are arranged in the same way as in the answer column in TABLE I.

### D. The Effect of the Previous Questions

For each question in the sessions, it is checked that this question requires which of the previous questions to be completed. In Fig. 1, it can be seen that more than half of the questions are related to the first question of the session and then the questions need the previous question to be completed. Accordingly, the previous question and the first question is included in the retrieval of related documents. The user's last question is arguably the most influential in finding answers. According to the diagram in Fig. 1, after the last question, the first question of the session and then the previous question is most effective. It can be seen that the first question of the session has an effect of 3.25 times compared to the previous question, so the first question was included in the retrieval of related texts with a weight of 3.25 compared to the previous question. Assigning weights to the previous questions made the content of the user's request more complete. Finally, the relevant documents were retrieved using LM.

---
[1] https://github.com/omkia/IKT2020

TABLE I. DIFFERENT CATEGORIES OF QUESTIONS AND THEIR TYPE OF ANSWERS

| Question Type | Expected answer |
|---|---|
| (Who …?) | People |
| (Where …?) | Geographical Information |
| Class | Groups |
| (How much …?) | Numbers or Numerical Terms |
| (When …?) | Numbers that come with time or time expressions (like BC) |
| (Describe …) | Definitions or descriptions |
| (How …?) | The process of doing something |
| (Why …?) | Cause, usually comes with the word because |

## IV. LAUNCHING THE EXPERIMENT

### A. Data Set

TREC CAsT Conversational Assistant Track is one of the first attempts to build CS systems. The proposed method is evaluated using CAsT. The 2019 TREC CAsT dataset includes 30 training topics and 50 assessment topics. Each topic consists of a sequence of queries. Out of 30 educational subjects, only 13 subjects had at least one question with a relevant document. This repository contains three sets of data: MS MARCO [25], TREC CAR [26] (including Wikipedia texts), and the Washington Post [15] (including the text of news articles), which there is about a total of 38 million texts. In TABLE II, you can see a sample session. Each session includes the title and the description of the user's information needs. The session begins with a question on the subject and continues with other questions. *N* is equal to 80 in total. In this user session, since we have 12 questions, *t* is equal to 12. One thousand documents are extracted for each question, so *M* is equal to one thousand.

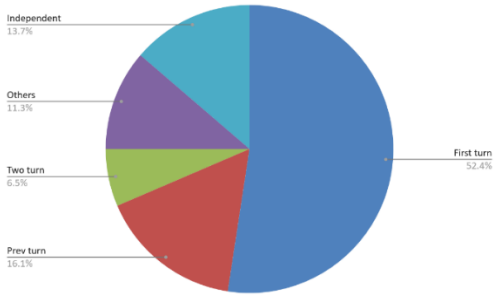

Fig. 1. Chart of the dependence of the current question on the previous questions

TABLE II. TABLE 1: SERIES OF QUESTIONS RELATED TO JOB SELECTION IN TREC CAsT

| Title: | career choice for Nursing and Physician's Assistant |
|---|---|
| Description: | Considering career options for becoming a physician's assistant vs. a nurse. Discussion topics include required education (including time, cost), salaries, and which is better overall |
| 1 | What is a physician's assistant? |
| 2 | What are the educational requirements required to become one? |
| 3 | What does it cost? |
| … | … |
| 11 | What is the fastest way to become an NP? |
| 12 | How much longer does it take to become a doctor after being an NP? |

### B. Baseline Method

In the baseline CAsT method, the preprocessing presented in Section III-B is applied. After preprocessing, the results were extracted using Query Likelihood (QL). This is the most common method of initial information retrieval used in neural networks [10], [11].

### C. Evaluation Criteria

In neural networks, reclassification is usually performed on an initial set of 1000 documents, so the Recall @ 1000 criterion is examined in Fig. 2. Also, the performance of the system in the case of being used without applying neural networks is measured using the NDCG criterion.

### D. Parameter Configuration

In this research, LM is used along with Dirichlet smoothing in which *µ = 2500*. In the proposed method of this article, two steps of classifying questions and influencing previous questions have been applied. The effect of the first question of the session is 3.25 times that of the previous question in each session.

## V. RESULTS

In the Recall @ 1000 criterion, by applying the classification and considering the context of the questions, the average improvement was 22% compared to the average of other methods, for example, it was improved from about 52% in the baseline CAsT method to 74%. Also in the NDCG @ 1000 criterion, the results have improved by an average of 10%. According to Fig. 3, the best improvement has been made in the category of questions related to *when*, but most of the questions are of description type, and therefore the dependence of the overall result on this part is observed.

## VI. CONCLUSION

In the dataset used in this study, it is not clear which part of the text is related to the question and the number of related questions and answers is extremely low. The method proposed in this article obtains more relevant answers in the retrieved documents than the common methods and also the accuracy of the results is improved. We will continue to complete and further analyze the dataset and explore ways to increase neural network training data. On the other hand, we will examine methods for estimating related sentences and will use BERT to extract the words or rules associated with each question.

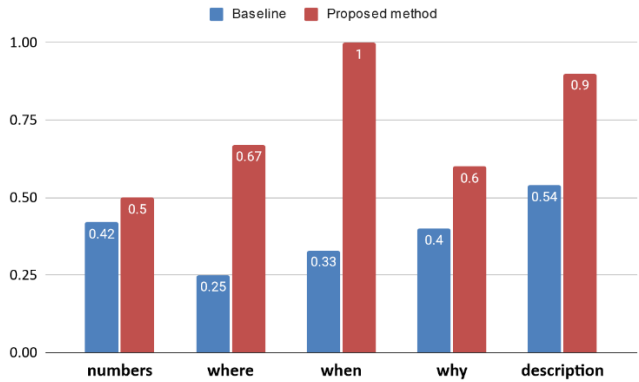

Fig. 2. Graph of Recall criteria improvement in each category

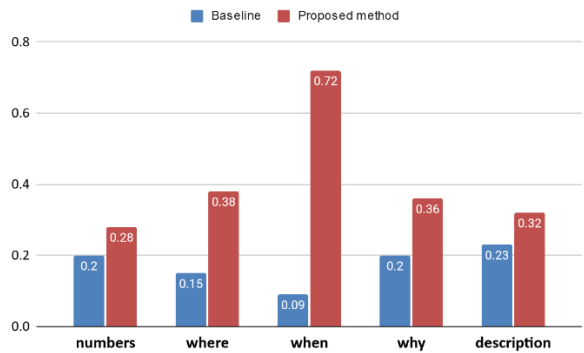

Fig. 3. Graph of NDCG criteria improvement in each category

REFERENCES

[1] A. Strzelecki, P. R.-M. and S. Technologies, and undefined 2020, "Featured snippets results in Google Web search: An exploratory study," *Springer*.

[2] H. Huang, K. Irene, and N. Ryu, "Voice Search and Typed Search Performance Comparison on Baidu Search System," Nov. 2019.

[3] J. Gao, M. Galley, and L. Li, "Neural approaches to conversational AI," in *41st International ACM SIGIR Conference on Research and Development in Information Retrieval, SIGIR 2018*, 2018, pp. 1371–1374, doi: 10.1145/3209978.3210183.

[4] S. Sahay, A. Venkatesh, A. R.-S. J. Delany, and undefined 2009, "Collaborative Information Access: A Conversational Search Approach," *pdfs.semanticscholar.org*.

[5] N. Voskarides, D. Li, A. Panteli, and P. Ren, "ILPS at TREC 2019 Conversational Assistant Track," 2019.

[6] S. Mehrotra and A. Yates, "MPII at TREC CAsT 2019: Incoporating Query Context into a BERT Re-ranker," 2019.

[7] J. Devlin, M.-W. Chang, K. Lee, and K. Toutanova, "BERT: Pre-training of Deep Bidirectional Transformers for Language Understanding," Oct. 2018.

[8] J. Xin, R. Tang, J. Lee, Y. Yu, J. L. preprint arXiv:2004.12993, and undefined 2020, "DeeBERT: Dynamic Early Exiting for Accelerating BERT Inference," *arxiv.org*.

[9] O. Khattab, M. Z. preprint arXiv:2004.12832, and undefined 2020, "ColBERT: Efficient and Effective Passage Search via Contextualized Late Interaction over BERT," *arxiv.org*.

[10] W. Yang, H. Zhang, and J. Lin, "Simple Applications of BERT for Ad Hoc Document Retrieval," Mar. 2019.

[11] Um. Amherst *et al.*, "Computer Science Department Faculty Publication Series," 2004.

[12] J. Guo *et al.*, "A Deep Look into neural ranking models for information retrieval," *Inf. Process. Manag.*, 2019, doi: 10.1016/j.ipm.2019.102067.

[13] W. Wu and R. Yan, "Deep Chit-Chat: Deep Learning for Chatbots," *dl.acm.org*, pp. 1413–1414, Jul. 2019, doi: 10.1145/3331184.3331388.

[14] Z. Zhang, R. Takanobu, Q. Zhu, M. Huang, and X. Zhu, "SCIENCE CHINA Technological Sciences. REVIEW. Recent Advances and Challenges in Task-oriented Dialog Systems," *Springer*, doi: 10.1007/s11432-016-0037-0.

[15] J. Dalton, C. Xiong, and J. Callan, "CAsT 2019: The Conversational Assistance Track Overview," 2020.

[16] L. Yang *et al.*, "Response ranking with deep matching networks and external knowledge in information-seeking conversation systems," in *41st International ACM SIGIR Conference on Research and Development in Information Retrieval, SIGIR 2018*, 2018, pp. 245–254, doi: 10.1145/3209978.3210011.

[17] C. Qu, L. Yang, M. Qiu, W. B. Croft, Y. Zhang, and M. Iyyer, "BERT with History Answer Embedding for Conversational Question Answering," 2019, pp. 1133–1136, doi: 10.1145/3331184.3331341.

[18] S. M. H. Nirob, M. K. Nayeem, and M. S. Islam, "Question classification using support vector machine with hybrid feature extraction method," in *20th International Conference of Computer and Information Technology, ICCIT 2017*, 2018, vol. 2018-January, pp. 1–6, doi: 10.1109/ICCITECHN.2017.8281790.

[19] D. Di Sarli, C. Gallicchio, and A. Micheli, "Question Classification with Untrained Recurrent Embeddings," in *Lecture Notes in Computer Science (including subseries Lecture Notes in Artificial Intelligence and Lecture Notes in Bioinformatics)*, 2019, vol. 11946 LNAI, pp. 362–375, doi: 10.1007/978-3-030-35166-3_26.

[20] T. Mikolov, K. Chen, G. Corrado, and J. Dean, "Efficient Estimation of Word Representations in Vector Space," Jan. 2013.

[21] M. Wasim, M. N. Asim, M. U. Ghani Khan, and W. Mahmood, "Multi-label biomedical question classification for lexical answer type prediction," *J. Biomed. Inform.*, vol. 93, p. 103143, May 2019, doi: 10.1016/j.jbi.2019.103143.

[22] T. Dodiya, S. J.-2016 I. I. W. C. on, and undefined 2016, "Question classification for medical domain question answering system," *ieeexplore.ieee.org*.

[23] K. Lee, L. He, M. Lewis, and L. Zettlemoyer, "End-to-end Neural Coreference Resolution," *EMNLP 2017 - Conf. Empir. Methods Nat. Lang. Process. Proc.*, pp. 188–197, Jul. 2017.

[24] M. Gardner *et al.*, "AllenNLP: A Deep Semantic Natural Language Processing Platform," pp. 1–6, Mar. 2018.

[25] P. Bajaj *et al.*, "MS MARCO: A Human Generated MAchine Reading COmprehension Dataset."

[26] L. Dietz, M. Verma, F. Radlinski, N. C.- TREC, and undefined 2017, "TREC Complex Answer Retrieval Overview.," *trec.nist.gov*.